\documentclass[12pt,psfig]{article}
\usepackage{amssymb}
\usepackage{amsmath,amsthm}
\usepackage{amsfonts}
\usepackage{graphicx}
\usepackage{graphics}
\usepackage{indentfirst}
\numberwithin{equation}{section}

\usepackage{hyperref}

\begin{document}
\begin{center}\Large\textbf{The Inflationary
Quartic Hilltop Model in a Modified Gravity
and Its Comparison with the Observations}
\end{center}
\vspace{0.75cm}
\begin{center}{\large Feyzollah Younesizadeh and \large Davoud
Kamani}
\end{center}
\begin{center}
\textsl{\small{Department of Physics, Amirkabir University of
Technology (Tehran Polytechnic) \\
P.O.Box: 15875-4413, Tehran, Iran \\
e-mails: fyounesizadeh@aut.ac.ir , kamani@aut.ac.ir \\}}
\end{center}
\vspace{0.5cm}

\begin{abstract}

We investigate the inflation for the quartic
hilltop model via a certain type of modified
gravity. Precisely, we analyze the
$F(\phi) T$ term in the Einstein's
gravity to examine the quartic hilltop inflation model.
$T$ is the trace of the energy-momentum tensor,
and $\phi$ is the inflaton field.
Next, we calculate the inflation
dynamics for the foregoing model
and obtain the slow-roll parameters, i.e., the scalar
spectral index ``$n_s$'' and the
tensor-to-scale ratio ``$r$'', which these parameters
exhibit high sensitivity to the $F(\phi) T$ term.
This modified form of the gravity is not only
in accordance with the
predictions of the original model but also allows for better
prediction of the Planck/BICEP/Keck data.

\end{abstract}


\newpage
\section{Introduction}

Research on cosmological microwave background (CMB)
shows a flat, evenly spread universe with accelerated expansion,
described by combining Einstein-Hilbert action with
inflaton action \cite{1}-\cite{4}.
The slow-roll inflation, where the inflaton
potential $V(\phi)$ drastically dominates
the energy density, involves the inflaton
field $\phi$ undergoing
a slow, nearly constant roll, and the potential
$V(\phi)$ remains nearly constant \cite{5}, \cite{6}.

The recent data from the Planck, BICEP, and Keck Array
\cite{7}-\cite{9} strongly favor concave inflationary
potentials over the convex ones. The hilltop models
\cite{10}-\cite{17} suggest a simple way to represent concave
potentials. All concave-shaped potentials can be approximated
by a hilltop potential. However, due
to the fact that the potential
is unbounded from below, some limitations exist. Squaring the
hilltop potential \cite{11}, \cite{15}
can generate a sombrero-hat
potential, as seen in the symmetry-breaking of the inflaton
model \cite{16}-\cite{18}.
The potentials $V = V_{0}(1-\frac{\phi^n}{m^n})+\cdots$
are frequently used for analyzing the hilltop inflation models.
The omitted terms, indicated by dots, do not influence the
inflation for $m\lesssim 1$. The models with $n = 2$ and $4$
for $m = 1$ are observationally ruled out. Hence, the
tensor-to-scalar ratio ``$r$'' computations in the models with
$V = V_{0}(1-\frac{\phi^n}{m^n})$, for large ``$m$''
and any $n$, fall into $r = 4/N$ limit.
In the hilltop models, the inflaton as an adequate field
independently moves of a state of instability.
This is the quartic hilltop model with $n=4$, which
we shall apply it.

Small-field inflationary models like hilltop inflation
have negative potential curvature and are better
understood in effective field theory with inflaton
field variance $\lesssim M_{P}$. These potentials
often stem from particle physics models with
symmetry breaking, making arguments over initial
conditions irrelevant due to everlasting inflation
in peak inflation \cite{19}.

The brane universe is a natural manifestation of the
nature in the string theory \cite{20}-\cite{25}.
A successful complement to the
$\alpha$-attractors \cite{26}-\cite{30} in
characterizing most of the regions
in the $(n_{s}, r)$-space, favored
by Planck 2018, reliably comes from the D-brane
inflation models. Besides, the inflation may originate from
other dimensions, as suggested by the D-brane inflation.
The power-law potentials \cite{31} are
popular inflationary models,
featuring a single monomial potential.
In these models, the inflation
occurs at high values of the inflaton field $\phi$.
The natural inflation
\cite{32}, \cite{33} relies on the periodic potentials.
They contain a pseudo-Nambu-Goldstone boson
as the inflaton. The other useful model
is the $R^2$-inflation, which was proposed by Starobinsky
\cite{34}, \cite{35}. The exponential tails models
\cite{36} have been carried out by the
supergravity and string theory.

A crucial approach in the inflationary process is the
formation of GR activity via the geometric components,
matter components, and or both.
Thus, the modified gravities
with $f(R)$, $f(T)$, $f(G)$, and $f(R, T)$ were appeared
\cite{37}–\cite{43}.
Hence, the term $f(\phi)T$ has been
added to the Einstein Hilbert
action \cite{44}, in which $T$ represents the trace of
the energy-momentum tensor.
In this case, the authors studied the inflationary
epoch via the different effective
potentials and various $f(\phi)$. These potentials
are the chaotic, natural, and
Starobinsky inflation in the conformal frame.
The idea of extending the gravity with the
term $f(\phi,T)$ revives the inflationary
models, e.g., the chaotic and natural inflations.
The foregoing models, which are incompatible
with the standard GR, has
been reconsidered due to the observational discrepancies.

Among the modified gravity theories, the $F(\phi)T$-gravity
has the non-minimal coupling
of the inflaton field $\phi$ with the trace of
the energy-momentum tensor``$T$''
\cite{44}-\cite{47}. This term is
generally motivated by the quantum gravity.
It also provides a description of the unified gravity
with other fields \cite{45}-\cite{48}.
The $F(\phi)T$-gravity is an
extension of the $f(R, T)$-gravity. The slow-roll
inflation in the presence of the simple term
$F(\phi)T = \sqrt{\kappa}\phi T$ has
been investigated \cite{44}. It revealed that the
Natural, Starobinsky and Chaotic models behave
in better agreement with the observational data.

In this paper, we consider a modified
gravity with the Lagrangian
density ${R}/{2\kappa}+\beta F(\phi)T+ \mathcal{L}_{m}(\phi)$.
At first, we start with a general
scalar potential $V(\phi)$ and an
arbitrary functional $F(\phi)$. Afterward, we apply
the quartic hilltop potential and a specific form
of $F(\phi)$. We compute the inflation
via the foregoing configuration.
We calculate the slow-roll parameters, i.e., the scalar
spectral index ``$n_s$'' and the tensor-to-scale ratio ``$r$''.
Our model is in accordance with the
predictions of the original model. Besides,
it also allows for better predictions
of the Planck/BICEP/Keck data.

The paper is organized as follows.
In Sec. \ref{200}, we calculate
the equations of motion, the slow-roll parameters,
``$n_s$'' and ``$r$'' in the presence of the term
$F(\phi) T$. In Sec. \ref{300}, we investigate
the influence of this term on the quartic hilltop
inflationary model. We present a comparative analysis
between the calculated data and the available data
from the Planck 2018. Sec. \ref{400} is entirely devoted to
the conclusions.

\section{The inflation via the $f(\phi,T)$ theory}
\label{200}

With the scalar field $\phi$ coupled with the
energy-momentum ($T$) tensor, the modified
gravity $f(\phi,T)$ model's action can be
expressed as follows \cite{44},
\footnote{We use the natural units
$c =\hbar = 1$, $\kappa=8\pi G$, and
the metric signature is $(-,+,+,+)$.}
\begin{equation}
\label{16}
S=\int{{\rm d}^4x\sqrt{-g}}\left(\frac{R}{2\kappa}+f(\phi,T)+
\mathcal{L}_{m}({\phi,\partial_{\mu}\phi})\right),
\end{equation}
where $f(\phi,T)$ denotes
a functional of $\phi$ and $T=g^{\mu\nu}T_{\mu\nu}$.
The Lagrangian density
$\mathcal{L}_{m}$ refers to the matter sector.

The universe, during the inflationary epoch, should
be primarily influenced by a solitary inflaton field $\phi(t)$.
This field obviously contributes to the matter
Lagrangian $\mathcal{L}_{m}$ as
\begin{equation}
\label{5}
\mathcal{L}_{m}({\phi,\partial_{\mu}\phi})
=-\frac{1}{2}g_{\mu\nu}\partial^{\mu}\phi
\partial^{\nu}\phi-V(\phi)
=\frac{1}{2}\dot \phi^2-V(\phi),
\end{equation}
where $V(\phi)$ represents the potential of the scalar field.
We assumed that the inflaton field is
spatially homogeneous and varies only with respect to
the time ``$t$''.

The energy-momentum tensor, extracted from the
above action, is given by
\begin{equation}
T_{\mu\nu}=g_{\mu\nu}\mathcal{L}_{m}-2\frac{\delta
\mathcal{L}_{m}}{\delta
g^{\mu \nu}}=\partial_{\mu}\phi\partial_{\nu}
\phi+g_{\mu\nu}\left(\frac{1}{2}\dot \phi^2-V(\phi)\right).
\end{equation}

If we consider the universe as a perfect fluid, the
pressure ``$p$'' and the energy density ``$\rho$'',
are prominently expressed as in the following
\begin{equation}
T_{ij}=\Big(\frac{1}{2}\dot \phi^2-V(\phi)\Big)g_{ij}=pg_{ij},
\end{equation}
\begin{equation}
T_{00}=\frac{1}{2}\dot \phi^2+V(\phi)=\rho ,
\end{equation}
\begin{equation}
T=g^{\mu\nu}T_{\mu\nu}=\dot \phi^2-4V(\phi).
\end{equation}

The action \eqref{16} elaborates the Einstein equation
\begin{equation}
\label{01}
R_{\mu\nu}-\frac{1}{2}g_{\mu\nu}R=\kappa \left(T_{\mu\nu}\big
[1-2f_{T}(\phi,T)\big]+\big[g_{\mu\nu}f(\phi,T)-2f_{T}(\phi,T)
\Theta_{\mu \nu}\big]\right).
\end{equation}
Here $f_{T}(\phi,T)=\frac{\partial f(\phi,T)}{\partial T}$ and
the symmetric tensor
$\Theta_{\mu \nu}\equiv g^{\xi \upsilon} \frac{\delta
T_{\xi \upsilon}}{\delta g^{\mu \nu}}$ possesses the feature
\begin{equation}
\label{2.8}
\Theta_{\mu \nu}\equiv g^{\xi \upsilon}
\frac{\delta T_{\xi \upsilon}}
{\delta g^{\mu \nu}}=-2T_{\mu \nu}+g_{\mu \nu}
\mathcal{L}_{m}-2g^{\xi \upsilon}
\frac{\delta^2 \mathcal{L}_{m}}{\delta
g^{\mu \nu}\delta g^{\xi \upsilon}}.
\end{equation}

The Lagrangian \eqref{5} gives the explicit form of
this tensor
\begin{equation}
\Theta_{\mu \nu}=-\partial_{\mu}\phi
\partial_{\nu}\phi-T_{\mu\nu}
=-2\partial_{\mu}\phi\partial_{\nu}\phi-g_{\mu\nu}
\left(\frac{1}{2}\dot \phi^2 -V(\phi)\right).
\end{equation}

From now on we apply the form $f(\phi,T)=\beta
F(\phi)T$, where $F(\phi)$ is an arbitrary functional
of $\phi$, and $\beta$ is a constant parameter. In the limit
$\beta\to$ 0, our model manifestly reduces to the
Einstein gravity. We shall see that fitting the inflationary
predictions of the model
with the observations only requires a very small
value of $\beta$. We should note that
our model, after the inflation, reverts to the
Einstein's gravity.

Now we rewrite Eq. \eqref{01} in the form
\begin{equation}
\label{2.10}
R_{\mu \nu}-\frac{1}{2}g_{\mu \nu}R =
\kappa T_{\mu \nu}^{({\rm eff})},
\end{equation}
\begin{equation}
\label{2.11}
T_{\mu \nu}^{({\rm eff})} = T_{\mu \nu}-2\beta
F\left(T_{\mu \nu}-\frac{1}{2}Tg_{\mu \nu}
+\Theta_{\mu \nu}\right).
\end{equation}
The effective energy-momentum tensor \eqref{2.11}
defines the following effective energy density and pressure
\begin{equation}
\label{2.12}
\rho_{{\rm eff}}=T_{00}^{({\rm eff})}
=\frac{1}{2}\dot \phi^2 (1+2\beta F)
+(1+4\beta F)V,
\end{equation}
\begin{equation}
\label{2.13}
p_{{\rm eff}}g_{ij}=T_{ij}^{({\rm eff})}
=\left[\frac{1}{2} \dot
\phi^2(1+2\beta F)-(1+4\beta F) V\right]g_{ij}.
\end{equation}

Combining the FRW metric with Eqs. \eqref{2.12} and \eqref{2.13}
conveniently yields the Friedmann equations
\begin{equation}
\label{2.14}
H^2=\frac{\kappa \rho_{\rm eff}}{3}= \frac{\kappa}{3}
\left[\frac{\dot \phi^2}{2}(1+2\beta F)
+(1+4\beta F)V\right],
\end{equation}
\begin{equation}
\frac{\ddot{a}}{a}=-\frac{\kappa}{6}(3p_{\rm eff}
+\rho_{\rm eff})=-\frac{\kappa}{3}
\left[\dot \phi^2(1+2\beta F)-(1+4\beta F)V\right],
\end{equation}
\begin{equation}
\label{2.16}
{\dot H}=\frac{\ddot{a}}{a}-H^2= -\frac{\kappa}{2}
\dot\phi^2\left(1+2\beta F\right).
\end{equation}

The continuity equation
${\dot \rho}_{\rm eff}+3H(\rho_{\rm eff}+p_{\rm eff})=0$
is obtained similar to the usual case.
Besides, the modified Klein-Gordon equation is acquired by
combining Eq. \eqref{2.16} and the
time derivative of Eq. \eqref{2.14},
\begin{equation}
\ddot \phi+3H\dot\phi(1+2\beta F)+\beta
F^{{\prime}}\dot\phi^2+(1+4\beta F)
V^{{\prime}}+4\beta F^{{\prime}}V=0,
\end{equation}
where $F^{\prime}=\frac{dF}{d\phi}$ and
$V^{\prime}=\frac{dV}{d\phi}$.
The conditions ${\dot \phi^2}\ll V$,
$\ddot \phi \ll H \dot \phi$,
and $F^{{\prime}}\dot\phi^2\ll H \dot \phi$ are induced by
the slow-roll techniques. Consequently, in these approximations,
the Friedman and the modified Klein-Gordon equations
take the following features
\begin{equation}
\label{2.18}
3H\dot\phi(1+2\beta F)+(1+4\beta F)V^{\prime} +4\beta
F^{\prime}V\simeq0,
\end{equation}
\begin{equation}
\label{2.19}
H^2\simeq\frac{\kappa}{3}(1+4\beta F)V.
\end{equation}

In the foregoing approximation, the slow-roll
parameters are defined by \cite{6},
\begin{equation}
\label{14}
\epsilon_{\rm E}=\frac{1}{2}M_{pl}^2\left(
\frac{V^{\prime}}{V}\right)^2, \qquad
\eta_{\rm E}=M_{pl}^2\frac{V^{\prime \prime}}{V},
\end{equation}
the subscribe ``E'' represents the Einstein's gravity. The
requirements for the slow-roll inflation
are $\epsilon_{\rm E}<1$ and $\vert\eta_{\rm E}\vert<1$.
In our model, the slow-roll parameters find the forms
\begin{equation}
\label{30}
\epsilon_V\simeq\frac{1}
{2\kappa \big(1+2\beta F\big)}
\left(\frac{V^{\prime}}{V}+\frac {4\beta F^{\prime}}{1+4\beta
F}\right)^2,
\end{equation}
\begin{equation}
\label{31}
\eta_V\simeq\frac{1}
{\kappa\big(1+2\beta F\big)}
\left[\frac{V^{\prime\prime}}{V}
+\frac{2\beta\big(3+4\beta F\big)
F^{{\prime}}}{\big(1+2\beta F\big) \big(1+4\beta F\big)}
\frac{V^{{\prime}}}{V}+\frac{4\beta
\big(1+2\beta F\big) F^{\prime\prime}-8\beta^2F^{\prime 2}}
{\big(1+2\beta F\big) \big(1+4\beta
F\big)}\right],
\end{equation}
where $F^{\prime\prime}=\frac{d^{2}F}{d\phi^{2}}$ and
$V^{\prime\prime}=\frac{d^{2}V}{d\phi^{2}}$.
In the limit $\beta \to 0$, these
parameters obviously reduce to ${\epsilon_{\rm E}}$ and
$\eta_{\rm E}$, as expected.
In fact, for fitting our predictions
with the observations, the value of the parameter
``$\beta$'' should be sufficiently small.
Thus, the linear order expansion is a very good approximation.

The number of e-folds, or $N =\ln(a)$, is
a quantity that reveals how far the spacetime
has been stretched. In our model, $N$ is given by
\begin{equation}
N\simeq\int_{\phi_i}^{\phi_f} {\rm d}\phi
\left(\frac{\kappa V}{V^{\prime}}
+ \frac{2\kappa\beta V\big(FV^{\prime}-2F^{\prime} V\big)}
{V^{\prime 2}}\right),
\end{equation}
where the upper (lower) limit $\phi_f$
($\phi_i$) is the inflaton value at the
horizon crossing (at the end of inflation).

Under the conditions of the slow-roll
approximation, the scalar power spectrum of the curvature
perturbations, $A_s$, and $n_s$, defined
by $n_s-1=\frac{d\ln A_s}{d\ln k}$, can be expressed in terms
of the slow-roll parameters as follows \cite{49},
\begin{equation}
\label{2.24}
A_s=\frac{3\kappa H^2}{24\pi^2 \epsilon_V}\simeq
\frac{\kappa^2(1+4\beta F)V}{24\pi^2 \epsilon_V},
\end{equation}
\begin{equation}
\label{2.25}
n_s=1+\frac{d\ln A_s}{d\ln k}\simeq 1+2\eta_V-6\epsilon_V.
\end{equation}
Note that the following constraint on $A_s$ is obtained from
the Planck, BICEP/Keck, and the
other observations \cite{50},
\begin{equation}
A_s=(2.10\pm0.03)\times10^{-9}.
\end{equation}

Using the slow roll parameters $\epsilon_V$ and $\eta_V$,
we can show that $n_s$ in Eq. \eqref{2.25}
can be obtained from Eq. \eqref{2.24},
\begin{align*}
\label{2.27}
n_s = 1+\frac{{\rm d}\ln A_s}{{\rm d}\ln k}
\simeq 1+\frac{{\rm d}\ln A_s}{{\rm d} N}=
1+\frac{{\rm d}\ln A_s}{H {\rm d}t}\\
= 1+\frac{\dot\phi}
{H A_s}\frac{{\rm d}A_s}{{\rm d}\phi}
=1+\frac{\dot\phi}{H}\Big[\frac{4\beta F^{\prime}}
{1+4\beta F}+\frac{V^{\prime}}{V}-\frac{\epsilon_V^{\prime}}
{\epsilon_V}\Big], \tag{2.27}
\end{align*}
the approximation ${\rm d} \ln k \simeq {\rm d}N$
has been employed \cite{6}.
By substituting Eqs. \eqref{2.19} and \eqref{30} into
Eq. \eqref{2.27}, we received Eq. \eqref{2.25}.

Regarding the production of tensor perturbations in the
inflationary period, the amplitudes of the gravitational
waves grow as follows \cite{49},
\begin{equation}
\label{2.28}
A_t=\frac{2\kappa}{\pi^2}H^2\simeq
\frac{2\kappa^2(1+4\beta F)V}{3\pi^2}.\tag{2.28}
\end{equation}

The relationship between the tensor to scalar ratio
``$r$'' , and the slow roll parameter $\epsilon_V$ is
established through Eq. \eqref{2.28}
and  \eqref{2.24} as follows
\begin{equation}
\label{2.29}
r = A_t/A_s \simeq16\epsilon_V.\tag{2.29}
\end{equation}

Concerning the tensor spectrum index $n_t$, which is
defined by $n_t = {\rm d}\ln A_t/{\rm d} \ln k$,
we can represent
this index alternatively as follows
\begin{align*}
\label{2.30}
n_t \simeq
\frac{{\rm d}\ln A_t}{{\rm d} N}=\frac{\dot\phi}{H}
\frac{{\rm d}\ln A_t}{{\rm d}\phi}=\frac{\dot\phi}{H A_t}
\frac{{\rm d}A_t}{{\rm d}\phi}
\\
= -\frac{1}{\kappa(1+2\beta F)}
\Big[\frac{4\beta F^{\prime}V+(1+4\beta F)
V^{\prime}}{(1+4\beta F)V}\Big]^2=-2\epsilon_V,\tag{2.30}
\end{align*}
where we employed Eq. \eqref{2.18},
\eqref{2.19}, \eqref{30}, and \eqref{2.28}.
Comparing Eq. \eqref{2.29} and
Eq. \eqref{2.30}, we acquire $n_t=-r/8$,
which is consistent with the results of the Refs.
\cite{6}, \cite{49}.

For calculating the inflationary properties,
the slow-roll approximation is commonly employed.
The potential $V(\phi)$ and the functional
$F(\phi)$ define the slow-roll constraints. Assuming
that the slow-roll parameters are small is equivalent
to using the slow-roll approximation
for the general relativistic models.
These models include the minimally
linked scalar fields \cite{42}.

One of the inflation's
achievements is that it offers a way
for the structure development,
which eventually arises from the tiny
density perturbations in the
early universe \cite{51}, \cite{52}.
The field value at the horizon crossing
can be found by backward counting after
finishing the slow-roll and
ensuring that at least 50 e-folds have been formed.
With the field value at the horizon crossing and
matching the slow-roll parameters, the
tensor-to-scalar ratio ``$r$'' and the
spectral index ``$n_{\rm s}$'' can be written as
\begin{equation}
\label{2.9}
r= 16\epsilon_{\rm E},
\qquad
n_{\rm s}=1+2\eta_{\rm E} -6 \epsilon_{\rm E}.\tag{2.31}
\end{equation}

Using the above quantities, the
data have been compared \cite{53}. Based
on the Planck 2018 data, the
parameters of the cosmic disturbances
need to satisfy the following observational
limits \cite{8}, \cite{54}.
\begin{equation}
n_s=0.9663\pm0.0041,\qquad r<0.065 ,\tag{2.32}
\end{equation}
with $68\%$ C.L and $95\%$ C.L, respectively.
We shall use reliable cosmic evidences
to evaluate the feasibility
of the $F(\phi)T$-gravity hypothesis.

\section{The quartic hilltop inflation and the slow-roll}
\label{300}

The hilltop models are theoretical models that
describe the behavior of particles at
high energies. In these adequate
models, certain terms in their
equations are often neglected, which can simplify
the calculations but may also affect their accuracy.
The neglected term can vary,
depending on the specific model
and the considered energy range.
For example, some models may neglect terms
that are related to the
interactions between the particles, while others may neglect
terms that are related to the curvature of the spacetime.

In the hilltop models, the
inflation occurs close to the maximum of the potential
or ``hilltop''. These models are appealing
for the slow-roll inflation. They
have been investigated from the point of view of
supersymmetry \cite{55}, \cite{56},
supergravity \cite{57}-\cite{59},
superstring theory models \cite{60}
and symmetry-breaking theories \cite{61}.

The inflaton potential
$V(\phi) = V_{0}\left(1-\frac{\phi^n}{m^n}\right)+\cdots$
defines the hilltop model,
where $V_{0}$, ``$m$'' and ``$n$'' are all positive
constants. The dots represent extra terms that are
irrelevant to the inflationary regime. It is
expected that the additional terms, denoted by dots, will
generate at least some of the appropriate
potentials \cite{10}, \cite{62}.
The most straightforward explanation is
that these terms represent the higher order
when the approximation $\frac{\phi^n}{m^n}$ is considered.
For $n = 4$, this potential accurately depicts the behavior of
the Coleman-Weinberg potential.
It can be demonstrated that for small field models \cite{63}
with $m \ge 1$, the inflation takes place at $\phi \ge m$, where
the higher order terms are insignificant.
As a result, some ambiguities in the potential's definition at
$\phi \sim m$ do not impact the inflationary predictions.
Hence, in various researches on this topic,
the simplest models are used for
computing ``$n_s$'' and ``$r$''.
For example, the Planck 2018 applies \cite{50},
\begin{equation}
\label{3.1}
V(\phi) = V_{0}\left(1-\frac{\phi^n}{m^n}\right).
\end{equation}

Based on the early mathematical approaches, the
spectral index of the curvature perturbation is \cite{5},
\begin{equation}
n_{s}=1-\frac{2}{N}\left(\frac{n-1}{n-2}\right),
\end{equation}
where $N$ represents the total
number of the e-folds of inflation.

The hilltop inflation better fits the Planck data.
It is represented by the green band in Fig. 8 of the
Planck 2018 inflation. We shall also reproduced it
in Fig. 1. The Eq. \eqref{3.1} manifestly yields the
following quartic hilltop inflation model, which we
apply it
\begin{equation}
\label{3.3}
V(\phi) = V_{0}\left(1-\frac{\phi^4}{m^4}\right).
\end{equation}

\newpage
\begin{center}
\begin{figure}
\centering
\includegraphics[width=14cm]{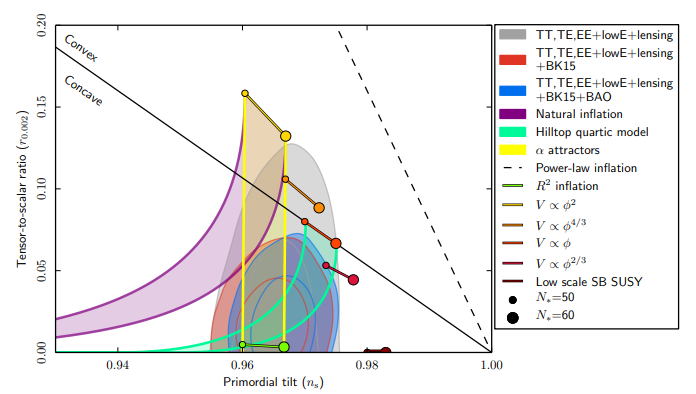}
\caption{\scriptsize{Using the Planck data
alone, BK15 data, and or BK15+BAO
data, the marginalized joint 68\% and
95\% CL areas for ``$n_{s}$''
and ``$r$'' at $k=0.002 Mpc^{-1}$ were
compared with the theoretical
predictions of specific inflationary theories. It should be
noted that ${\rm d}n_{s}/{\rm d}ln k = 0$ is assumed
in the combined 68\% and 95\% CL areas.}}
\end{figure}
\end{center}
Here, we put away the higher-order terms that stabilize the
potential. In other words, they become
significant only when the inflaton
achieves its vacuum expectation
value, but not during the inflation.
Our proposed functional $F(\phi)$ is as follows
\begin{equation}
\label{3.4}
F(\phi)=\frac{\phi}{\mu}\left[1+\lambda \ln
\left( \frac{\phi}{m}\right)\right].
\end{equation}
This form of $F(\phi)$
can be approximately considered as a modified
gravity with a non-minimal interaction between the gravity
and matter. In fact, we have extended the functional
$F(\phi)$ from its simple forms, e.g.
$F(\phi) = \phi$ \cite{44}, to a generalized form.
This functional satisfies the following expected limits
$F(0) = 0$, and $F(\phi) \simeq \phi$
for $\lambda \ll 1$ and $\phi/m\ll 1$.
However, this choice in the limit $\phi \to 0$
returns our setup to the Einstein's gravity.

Applying the quartic hilltop model conveniently yields
the explicit forms of the slow-roll parameters \eqref{14}
\begin{equation}
\epsilon_{\rm E}=\frac{8\left(\frac{\phi}{m}
\right)^6}{\left[1-\left(\frac{\phi}{m}\right)^4\right]^2}\;,
\end{equation}
\begin{equation}
\eta_{\rm E}=-\frac{12\left(\frac{\phi}{m}
\right)^2}{1-\left(\frac{\phi}{m}\right)^4}\;,
\end{equation}
where we employed $m = M_{\rm Pl}$.
The quartic potential \eqref{3.3}, the functional
\eqref{3.4} and the $F(\phi)T-$gravity
define the slow-roll parameters, i.e. Eqs.
\eqref{30} and \eqref{31}, as in the following
\begin{eqnarray}
\epsilon_V&\simeq&\frac{8\mu\Bigg( \beta\,
\lambda\, \left( {m}^{4}-5\,{\phi}^{4} \right) \ln
\left( {\frac {\phi}{m}} \right) - \beta
\Big[ \left( \lambda+5 \right)
{\phi}^{4}-{m}^{4} \left( \lambda+1 \right)
\Big]-\mu\,{\phi}^3 \Bigg)^{2}}{ \kappa\Big( {m}^{4}-{\phi}^{4}
\Big) ^{2} \Big( 4\beta\,\lambda\,\phi\ln  \left( {\frac {
\phi}{m}} \right)+4\,\beta\,\phi+\mu \Big)^2
\Big( 2\beta\,\lambda\,\phi
\ln  \left( {\frac {\phi}{m}} \right)
+2\,\beta\,\phi+\mu \Big)}\;,
\end{eqnarray}
\begin{eqnarray}
\eta_V&\simeq&\frac{1}{\kappa\Big( {m}^{4}-{\phi}^{4}
\Big)\Big( 4\,\beta\,\lambda\,\phi \ln \left( {\frac {\phi
}{m}} \right) +4\,\beta\,\phi+\mu \Big) \Big(
2\,\beta\,\lambda\,\phi
\ln  \left( {\frac {\phi}{m}} \right)
+2\,\beta \,\phi+\mu \Big) ^{2}\phi\,
}\nonumber\\
&\times& 4\mu\Bigg(-2\,\phi\,{\beta}^{2}{\lambda}^{2}
\left( {m}^{4}+15\,{\phi}^{4}
\right) \ln^{2} \left( {\frac {\phi}{m}}\right)
\nonumber\\
&-&2\,\beta\, \left( 3\,\beta\,
\left( \lambda+10 \right) {\phi}^{4}+12
\,\mu\,{\phi}^{3}+{m}^{4}\beta\,
\left( \lambda+2 \right)  \right)
\phi\,\lambda\,\ln  \left( {\frac {\phi}{m}} \right)
\nonumber\\&+&2\,{\beta}^{2} \left( {\lambda}^{2}
-3\,\lambda-15 \right) {\phi}^{5}-7
\, \left( \lambda+{24/7} \right) \beta\,\mu\,{\phi}^{4}
\nonumber\\
&-&3\,{\mu}^{2}{\phi}^{3}-2\,{m}^{4}{\beta}^{2}
\left( {\lambda}^{2}+
\lambda+1 \right) \phi+\beta\,\lambda\,{m}^{4}\mu\Bigg).
\end{eqnarray}
These slow-roll parameters clearly return to
$\epsilon_{\rm E}$ and $\eta_{\rm E}$ as $\beta \to 0$.
It is no longer possible to define the slow-roll parameters
in terms of the e-folds $N$. Besides, it is not
possible to obtain an expression of the field value
at the end of the inflation, because it is
$\beta$-dependent.

Now we can compute the tensor-to-scalar ratio
and spectral index
\begin{eqnarray}
r&=& \frac{8\Bigg(-\frac{4\,{\phi}^{3}
}{{m}^{4} \left( 1-{\frac {{\phi}^{4}}{{m}^{4}}} \right)
}+\frac{4\,\beta\, \left( {\frac {1}{\mu}
\big[ 1+\lambda\,\ln  \left( {
\frac {\phi}{m}} \right)  \big] }
+{\frac {\lambda}{\mu}} \right)
}{1+4\,{\frac {\beta\,\phi}{\mu}
\big[ 1+\lambda\,\ln  \left( {\frac {
\phi}{m}} \right)  \big] }
}\Bigg)^2}{\kappa\, \left ( 1+2\,
{\frac {\beta\,\phi}{\mu} \big[ 1+\lambda\,\ln
\left( {\frac {\phi}{m}} \right)  \big] } \right) }
\end{eqnarray}
\begin{eqnarray}
n_{s}&=&1+\frac{1}{\kappa\,
\left[ 1+2\,{\frac {\beta\,\phi}{\mu} \big( 1+\lambda\,\ln
\left( {\frac {\phi}{m}} \right)  \big) } \right]
}\Bigg\{2\Bigg(-\frac{12\,{\phi}^{2}
}{{m}^{4} \left( 1-{\frac {{\phi}^{4}}{{m}^{4}}} \right)
}\nonumber\\
&-&\frac{8\,\beta\,
\left( 3+4\,{\frac {\beta\,\phi}{\mu} \big[ 1+\lambda\,
\ln  \left( {\frac {\phi}{m}} \right)
\big] } \right)  \left( {
\frac {1}{\mu} \big[ 1+\lambda\,\ln
\left( {\frac {\phi}{m}}
\right)\big] }+{\frac {\lambda}{\mu}} \right) {\phi}^{3}
}{\left( 1+4\,{\frac {\beta\,\phi}{\mu}
\big[ 1+\lambda\,\ln  \left( {
\frac {\phi}{m}} \right)  \big] } \right)
\left( 1+2\,{\frac {\beta
\,\phi}{\mu} \big[ 1+\lambda\,\ln
\left( {\frac {\phi}{m}} \right)
\big] } \right) {m}^{4}
\left( 1-{\frac {{\phi}^{4}}{{m}^{4}}} \right)}
\nonumber\\
&+&\frac{4\,{\frac {\beta\,
\lambda}{\mu\,\phi} \left( 1+2\,{\frac {\beta\,\phi
}{\mu} \big[ 1+\lambda\,\ln  \left( {\frac {\phi}{m}} \right)
\big] } \right) }-8\,{\beta}^{2}
\left( {\frac {1}{\mu} \big[ 1+
\lambda\,\ln  \left( {\frac {\phi}{m}}
\right)  \big] }+{\frac {\lambda}{\mu}} \right) ^{2}
}{\left( 1+4\,{\frac {\beta\,\phi}{\mu}
\big[ 1+\lambda\,\ln  \left( {
\frac {\phi}{m}} \right)  \big] } \right)
\left( 1+2\,{\frac {\beta
\,\phi}{\mu} \big[ 1+\lambda\,\ln
\left( {\frac {\phi}{m}} \right)
 \big] } \right) }\Bigg)
\nonumber\\
&-& 3\Bigg[-\frac{4\,{\phi}^{3}
}{{m}^{4} \left( 1-{\frac {{\phi}^{4}}{{m}^{4}}} \right)
}+\frac{4\,\beta\, \left( {\frac {1}{\mu}
\big[ 1+\lambda\,\ln  \left( {
\frac {\phi}{m}} \right)  \big] }+{\frac
{\lambda}{\mu}} \right)
}{1+4\,{\frac {\beta\,\phi}{\mu} \big[ 1+
\lambda\,\ln  \left( {\frac {
\phi}{m}} \right)  \big] }}\Bigg]^2 \Bigg\}.
\end{eqnarray}

The most common hilltop models with $n = 4$ depicted
in Fig. 1 predict $n_{s}=0.94$ for $N = 50$
and $n_{s}=0.95$ for $N = 60$.
Nevertheless, for $m\lesssim 1$, such
models are ruled out by observational evidence.
The Planck 2018 only gives the safer value
$n_{s}=0.96$, which these forecasts
approach for $m\gtrsim 10$. The green
lines in Fig. 1, which depict the
model predictions, converge at
the red circles. They represent the inflation predictions in
the proposed theory with a linear
potential $V\sim\phi$, at the large limit of
``$m$''. Besides, for all hilltop
potentials \eqref{3.1} as well
as for $n = 4$, a similar conclusion
has been acquired \cite{64}.

Now we arrange the tables 1 and 2 with $N = 50$ and 60,
respectively. Besides, we use
$\beta=0.02$, and ``$\mu$'' varies between the numbers 3.6 and
2.6, which in the figure 2 is displayed in the yellow color.
\begin{table}[!h]
\begin{center}
\caption{}
\label{demo-table}
\scriptsize{\begin{tabular}{||c c c c c c c c c c c c|}
\hline
$N$ & 50 & 50 & 50 & 50 & 50 & 50 & 50 &50 & 50 &50 &50\\
\hline
$\beta$  &  0.02 &   0.02 &   0.02 &   0.02 &  0.02
& 0.02 & 0.02 &  0.02&   0.02 &  0.02& 0.02\\
\hline
$\lambda$ & 4 & 4 &4 & 4 & 4 & 4 & 4 & 4& 4& 4 &4 \\
\hline
$\mu$ & 3.6 & 3.5 & 3.4 & 3.3 & 3.2 & 3.1 &3 &2.9&2.8&2.7&2.6 \\
\hline
$m$ & 10 & 10 & 10 & 10 & 10 & 10 &10&10&10&10&10 \\
\hline
$r$  & 0.00011 &0.00031  & 0.00062 & 0.00109&0.00173
&0.00260& 0.00375&0.00524&0.00717&0.00966&0.01284\\
\hline
$n_s$  & 0.94920 & 0.94974 & 0.95028 & 0.95081
&0.95132&0.95180 &0.95223&0.95260&0.95288&0.95304&0.95305 \\
\hline
\end{tabular}}
\end{center}
\end{table}

\newpage
\begin{table}[!h]
\begin{center}
\caption{}
\label{demo-table}
\scriptsize{\begin{tabular}{||c c c c c c c c c c c c|}
\hline
$N$ & 60 & 60 & 60 & 60 & 60  & 60 & 60 & 60 & 60 & 60 & 60\\
\hline
$\beta$  &  0.02 &   0.02 &   0.02 &   0.02 &  0.02
& 0.02 & 0.02 &  0.02&   0.02 &  0.02& 0.02\\
\hline
$\lambda$ & 4 & 4 &4 & 4 & 4 & 4 & 4 & 4& 4& 4 &4 \\
\hline
$\mu$ & 3.6 & 3.5 & 3.4 & 3.3 & 3.2 & 3.1 &3 &2.9&2.8&2.7&2.6 \\
\hline
$m$ & 10 & 10 & 10 & 10 & 10 & 10 &10&10&10&10&10 \\
\hline
$r$  & 0.00079 &0.00119  & 0.00171 & 0.00238&0.00323
&0.00431& 0.00568&0.00739&0.00955&0.01226&0.01567\\
\hline
$n_s$  & 0.96690 & 0.96757 & 0.96825 & 0.96894
&0.96964&0.97033 &0.97100&0.97165&0.97225&0.97276&0.97317 \\
\hline
\end{tabular}}
\end{center}
\end{table}

For the smallness of the parameter $\beta$, we should note
that we need an interval
for the values of $\beta$ to give us the best fit with
the data. Also, we added a modified term
to the action as a small term, in comparison with the
standard terms in the Einstein-Hilbert action, to
check its impact on the inflationary parameters.

In addition to the above tables, we changed ``$\mu$'',
and $\beta = 0.002$ was applied.
This case is shown in the figure 2 with the red color.

\newpage
\begin{center}
\begin{figure}
\centering
\includegraphics[width=17cm]{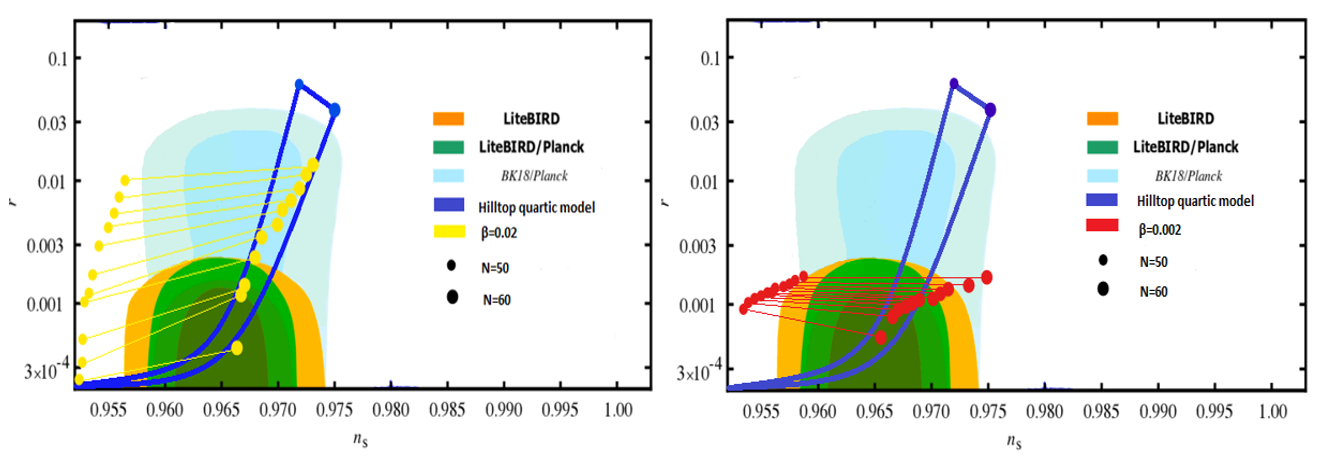}
\caption{\scriptsize{The index ``$n_{s}$'' and the ratio
``$r$'' in the modified gravity are predicted
by the hilltop inflation
theories. For comparison, the Einstein's gravity result
(see indigo bar) has been shown. In the present case,
the red colors for $\beta= 0.002$, $m =5$, $\lambda=3$
with $2.5 \leq \mu \leq 1.5$, and the
yellow colors for $\beta= 0.02$, $m =10$, $\lambda=4$ with
$3.6 \leq \mu \leq 2.6$ have been used in our model.
The parameters
``$n_{s}$'' and ``$r$'' at $k=0.002 Mpc^{-1}$ have C.L. margins
that are the dark (light) blue area
$1\sigma$ ($2\sigma$) region, respectively. The Planck
2018 data \cite{8} comprises the Einstein's gravity data.}}
\end{figure}
\end{center}

The indigo area of the Planck/BICEP/Keck data is incorporated
into the quartic hilltop inflation,
with a single-field model of
the Einstein's gravity, in figure 2.
By choosing suitable values for the parameters
``$\beta$'', ``$\mu$", ``$\lambda$'' and ``$m$'',
the term $F(\phi)T$ enabled us
to cover the region in the $(n_s, r)$-space that is favored by
the Planck 2018. Additionally, some
portions of the Planck data have been covered by using a
unique value for ``$\beta$'' but ``$\mu$'' as a variable. The
yellow and red colors in the
figure 2 indicate that by changing
the mentioned parameters, a large part of the Planck data
can be covered. In other words, the figure 2 shows how the
Planck data region can be approximately covered by
changing the parameters ``$\mu$'', ``$\lambda$'' and ``$m$''
and using two different values for ``$\beta$''.

The effects of the changes of the parameters
$\beta$, $\lambda$ and
$\mu$ on $n_s$ and $r$ have been illustrated in the figures 3
and 4. These diagrams represent
additional visual aids, such as the 
parameter sensitivity plots. The ranges of these
parameters that satisfy the Planck and BICEP/Keck
observational constraints have been shown.
The changes of $n_s$ and $r$ via the
parameter-changing provide an insight into the
robustness of the model.

\begin{center}
\begin{figure}
\centering
\includegraphics[width=18cm]{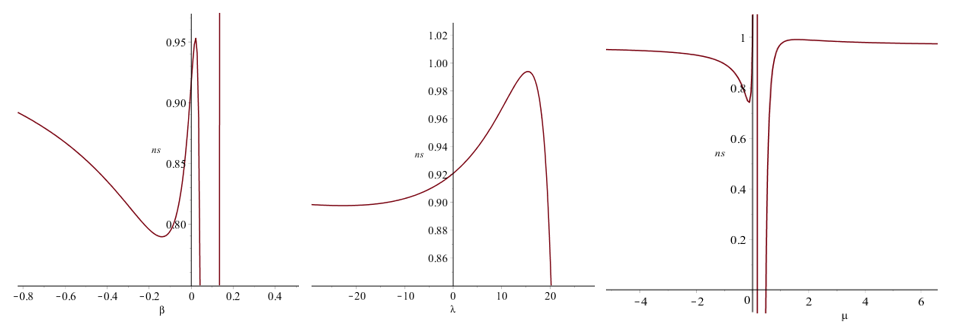}
\caption{\scriptsize{The sensitivity diagrams
of $n_s$ in terms of the parameters $\beta$,
$\lambda$ and $\mu$.}}
\end{figure}
\end{center}

\begin{center}
\begin{figure}
\centering
\includegraphics[width=18cm]{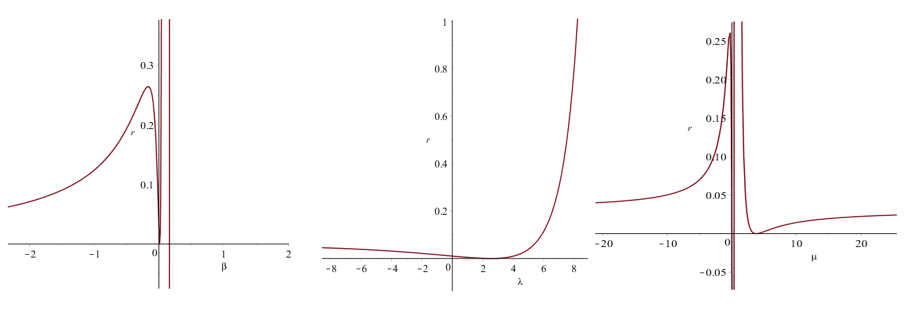}
\caption{\scriptsize{The sensitivity diagrams
of $r$ in terms of the parameters $\beta$,
$\lambda$ and $\mu$.}}
\end{figure}
\end{center}

Our model can be compared with the other inflationary
models, such as those with $F(\phi)=\phi$,
which highlight the trade-off between the
simplicity and predictability in the inflationary
dynamics. Although the fine-tuning may reduce the
predictive power, however, it can reflect the
imposed constraints by the observational
data (e.g., the Planck's limits on $n_s$ and $r$).
Thus, when one tries to reconcile
the theoretical predictions with the experimental
results the fine-tuning is unavoidable.

\section{Conclusions and discussion}
\label{400}

Revised concepts in the field of
the gravity always open a wide
range of theoretical explorations.
This expansion provides explanations for the astrophysical
compact objects. With this motivation, we investigated
a hilltop model within the framework of
$F(\phi) T$ gravity. The quartic hilltop model
is a desirable model that we examined it
in the gravity $F(\phi) T$.
Both scalar spectral index ``$n_{s}$'' and
tensor-to-scalar ratio ``$r$''
are drastically sensitive to this coupled
term. Using the cosmological observations
like ``$n_{s}$'' and ``$r$'', we selected
$F(\phi)$ in the form \eqref{3.4}.
Thus, both ``$n_{s}$'' and ``$r$''
are compatible with the Planck 2018.
Their corresponding
$(n_s, r)$-space falls within the
$2\sigma$ range and encompasses a larger section
of the $(n_s, r)$-space. We found that by choosing
appropriate values for the
parameters, the quartic hilltop model agrees with the
Planck's 2018. Besides, their
$(n_s, r)$-space enters the $2\sigma$ region
and covers a wider part of the $(n_s, r)$ plot.

The hilltop potentials need scales larger
than the Planck scale ($m\gg 10\;M_{\rm P}$) to
satisfy the current observational limits.
In accordance with the Planck 2018,
we examined this problem and concluded
that the scale ``$m$'' can be actually decreased
to less than ``$10\;M_{\rm P}$''. We presented
our results for the quartic hilltop potential
at the scales $10\;M_{\rm P}$ and $5 M_{\rm P}$.
In the figure 2, the observational results in
two cases are also exhibited.
In the first case, represented by the yellow
color in the figure 2, the values of the parameters
are $\beta=0.02$, $m=10$,
$\lambda=4$, and $3.6 \leq \mu \leq 2.6$.
Therefore, the data persist as yellow color in the
$(n_s, r)$-space plot within the specified
region of the Planck data. In the second case,
for the red portion, the parameters
are $\beta=0.002$, $m=5$, $\lambda=3$, and
$2.5 \leq \mu \leq 1.5$. The red points in
the $(n_s, r)$ plot remain in the $2\sigma$ region
and can actually cover the $1\sigma$
and $2\sigma$ regions in a more concentrated manner.
It can be seen that for the smaller values of $\mu$,
a larger area can be covered.
Note that in both above cases in the figure 2,
fine-tuning of ``$\lambda$'', ``$\mu$'',
and ``$\beta$'' can decrease the scale of
``$m$'' via the hilltop models to satisfy
the current observational constraints.

{\bf Discussion}

Our choice of the functional $F(\phi)$ significantly extends
the simpler models and aligns the theoretical
predictions with the experimental data. It may
also predict unique observational signatures,
like the primordial non-Gaussianity deviations or
power spectrum features. However, exploring
ways to reduce the fine-tuning and to enhance the
model's robustness would make it more
compelling and broadly applicable.
The role of the fine-tuning in our model is evident.
Future works could reveal the functional forms, and
also could display the theoretical
consistency with the observational data.

Our model is flexible. However, it contains the parameters
$\beta$, $\lambda$, $\mu$ and $m$ that may
complicate the outcome fitting with the
experimental data. Besides, the logarithmic term in $F(\phi)$
introduces a non-linear behavior that might lead
to theoretical instabilities or singularities, particularly
for the small values of $\phi$ or $\phi/m\approx 1$. Such
instabilities can challenge the robustness of the inflationary
predictions and may require a fine-tuning of the
parameters to avoid the non-physical solutions.


\end{document}